**Reward and adversity processing circuits, their competition and interactions with dopamine and serotonin signaling.**

Karin Vadovičová, Roberto Gasparotti
Neuroradiology Unit, Department of Diagnostic Imaging, Spedali Civili, 25123 Brescia, Italy.
email: vadovick@tcd.ie

**Abstract**
We propose that dorsal anterior cingulate cortex (dACC), anterior insula (AI) and adjacent caudolateral orbitofrontal cortex (clOFC), project to lateral habenula (LHb) and D2 loop of ventral striatum (VS), forming a functional adversity processing circuit, directed toward inhibitory avoidance and self-control. This circuit learns what is bad or harmful to us, evaluates and predicts risks - to stop us from selecting and going/moving for the bad or suboptimal choices that decrease our well-being and survival chances.
Proposed role of dACC is to generate a WARNING signal when things are going (or might end) bad or wrong, to prevent negative consequences: pain, harm, loss or failure. The AI signals about bad, low, noxious and aversive qualities, which might make us sick or cause discomfort.
These cortical adversity processing regions activate directly and indirectly (via D2 loop of VS) the LHb, which then inhibits dopamine and serotonin release (and is reciprocally inhibited by VTA/SNc, DRN and MRN) to avoid choosing and doing things leading to harm or loss, but also to make us feel worse, even down when overstimulated. We propose that dopamine attenuates output of the adversity processing circuit, thus decreasing inhibitory avoidance and self-control, while serotonin attenuates dACC, AI, clOFC, D1 loop of VS, LHb, amygdala and pain pathway.
Thus, by reciprocal inhibition, by causing dopamine and serotonin suppression - and by being suppressed by them, the adversity processing circuit competes with reward processing circuit for control of choice behavior and affective states. We propose stimulating effect of dopamine and calming inhibitory effect of serotonin on the active avoidance circuit involving amygdala, linked to threat processing, anger, fear, self-defense and violence. We describe causes and roles of dopamine and serotonin signaling in health and in mental dysfunctions. We add new idea on ventral ACC role in signaling that we are doing well and in inducing serotonin, when we gain/reach safety, comfort, valuable resources (social or biological rewards), affection and achieve goals.

**Adversity processing neural circuit - directed toward self-control, inhibitory avoidance and de-selection of harmful and suboptimal choices.**

We will characterize properties of the adversity processing circuit (APC), causally affecting value-based decision making, affective control of choice behavior, few mental disorders and well-being. We claim that the APC is in competition with the reward processing circuit (RPC) in guiding goal-directed behavior because APC directly and indirectly attenuates and competes with RPC - via its reciprocal connections, VS projections and suppression of dopamine and serotonin release. Activation of APC causes inhibitory avoidance and de-selection of bad, harmful or suboptimal choices. By suppressing the reward circuit, plus dopamine and serotonin signaling - the APC biases our thoughts, decisions, actions, emotions, motivation and learning.
Proposed circuit includes dACC, AI, adjacent clOFC, D2 loop of VS, GPe, subtalamic nucleus, ventral GPi and LHb. AI and dACC are reciprocally connected, form part of the pain pathway at its top - evaluative processing level, and are commonly co-activated in the NoGo task (inhibitory control task where the pre-potent response must get inhibited withdrawn at the occurrence of infrequent cue).

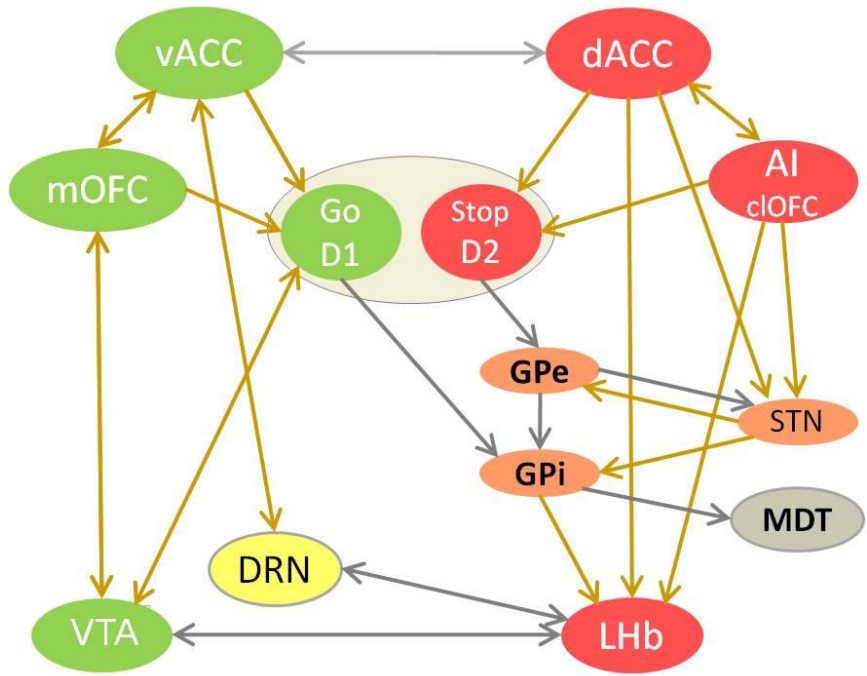

**Figure 1. The competition between reward and adversity processing circuit in choice behaviour.** Our model of value-based learning shows how the affective processing causes selection of good/valuable and de-selection of bad/harmful choices, by controlling dopamine and serotonin signaling in the brain. The implicit bias/inclination for the 'Go for it' versus 'Stop yourself - avoid it' response is learned in the motivational ventral striatum (VS), through the potentiation of cortical glutamatergic synapses by dopamine increase (at D1 loop) or its decrease (at D2 loop). Reward and adversity processing circuits are marked green and red. D1 and D2 neurons in VS are intermixed. The D1 neurons of VS disinhibit dopamine neurons by inhibiting the GABA interneurons in VTA. The indirect D2 loop (in orange) biases the choice selection toward inhibitory avoidance. The prefrontal projections of mediodorsal thalamus (MDT) enforce the representations of choices and goals in working memory, as part of cortico-striato-thalamo-cortical loop. Dopamine source VTA is marked green, serotonin source DRN yellow. The projections with excitatory effects are brown, with inhibitory effect are grey. Dopamine attenuates the output of the adversity processing circuit (in red) and potentiates that of the reward processing circuit (in green). Serotonin attenuates the AI, dACC, LHb, GPi, SNr, STN and motivational D1 loop of VS and enforces vACC, SNc (via SNr), GPe.

**Dorsal anterior cingulate cortex.**

Dorsal ACC is involved in both willful and unconscious control of goal-directed behavior. Many studies demonstrated its activation by performance monitoring, negative feedback and error detection – implying [1, 2] it detects conditions under which errors are likely to occur. dACC responds to pain and to attention demanding tasks [3], during response conflict and inhibition of pre-potent responses [4] and it activates the sympathetic system, heart rate and autonomic response to physical and mental stress [5-7]. The ACC was divided [8] based on its connectivity and control of goal-directed behavior, into cognitive dorsal (BA 24', 32') and affective ventral (subcallosal) region (BA 25, 24, 33), with pregenual ACC in between. dACC was linked to premotor response selection, noxious stimuli, effortful processing and motivation, while vACC was linked to rest and parasympathetic activation [8]. We summed up what the numerous observed dACC activations have in common. It seems to identify danger, harm, potential risks and bad or suboptimal outcomes of our actions, decisions and predictions.

We propose that the main computational role of dACC is to generate a WARNING signal (Beware! Attention!) toward the real or predicted bad outcomes - to call for changes in the current course of actions and behaviour, to prevent pain, harm, loss or failure. Also to prod for re-thinking and updating the wrong or faulty strategies, predictions, guesses, thus the cognitive models of the world around. This warning signal in the dACC is issued when things are going bad for us - leading to negative consequences: to increase our precaution, carefulness, engagement with the problem, attentional focus and cognitive-motor mobilization - to improve our performance and adjust the behaviour. Therefore the purpose of the dACC is to learn and predict the bad consequences of current actions and decisions, to warn us about harms, risks and failures (errors in actions, reasoning, thoughts, plans) to prevent bodily harm and personal loss. It also urges us to update/adjust the wrong, suboptimal or inappropriate behavior, strategies, plans - when getting negative feedback. Thus, the dACC detects danger and potential risk to prevent bad consequences and harm incurred by current actions, decisions and by behavior yet to be conducted. The proposed warning signal, generated by the dACC, resembles the negative surprise or unexpected uncertainty signal [9] found to drive noradrenaline release. But we claim that the dACC warning signal is also issued in response to rewarding or attractive stimuli, if there is a need to react quickly – not to miss/lose the passing chance of gain (i.e. prodding fast reaction to passing pretty woman). The dACC has all the connectivity needed - not only to identify the risky options and inadequate courses of actions to learn to avoid them, but also to react to potential danger, by recruiting the alarm response of brain and body. dACC passes its warning signal via top-down projections to several effector regions (Figure 2): to preSMA - to make a fast switch of current motor action, to STN - to stop inaccurate motor response, to frontal eye field and intraparietal cortex - to focus our visual and spatial attention, to noradrenergic and cholinergic nuclei to speed up reactivity and increase alertness, to the sympathetic system to increase heart and breath rate in demanding situations, plus to striatum, LHb and MDT to bias choice behavior. dACC projects to the stress-related hypothalamic nucleus and reciprocally to the threat processing amygdala. It provokes, by its robust reciprocal projections with lateral PFC and lateral BA 10, an update in failed predictions, decisions or strategies - to switch away from wrong/inadequate behavior to more optimal choices, actions or performance. The lateral BA 10 role is to seek to find out more fitting model, prediction or solution for what is going on, that can be applied to reach our goals and avoid harm. It generates novel guesses, hypotheses and ideas about studied systems by figuring out relevant interrelations, extracting contingencies, invariances and rules (IF - THEN), and explaining incongruities in the facts. Testing the validity of new predictions involves the lateral BA 10, dACC and dorsal striatum (DS).

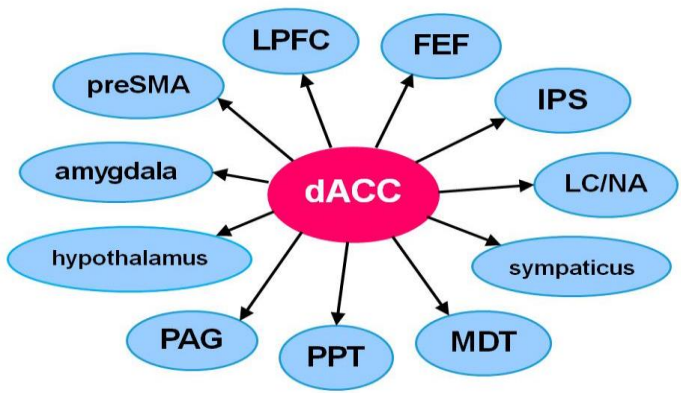

**Figure 2. A causal role of dACC connectivity in enabling propagation of its warning signal.**

The warning signal generated by dACC is forwarded to specific regions to induce caution, attention, arousal and cognitive-motor mobilization toward potential danger, to prevent suboptimal consequences. This signal is similar to 'Beware!' command and is evoked by demanding or risky situations and tasks. preSMA, presuplemmentary motor area, LPFC, lateral prefrontal cortex, FEF, frontal eye field, IPS, intraparietal sulcus, LC/NA ,locus coeruleus/noradrenaline, sympaticus, sympathetic system, MDT, mediodorsal thalamus, PPT, pedunculopontine nucleus, PAG, periaqueductal grey.

The more something matters to us (the more we value it), the stronger the warning signal (and worries) issued by dACC when we might lose it or fail (loss of resources, loved ones, social bonds, time, health, prospects, job). So the strength of dACC recruitment will depend on our subjective values and preferences. This warning signal is the possible source of error-related negativity (ERN), an event related potential [10] evoked by negative feedbacks and errors. An example for this dACC warning function is the study [11], which found that dACC stimulation elicits emotional vocalization in monkeys and humans, while its lesions affect voluntary control of emotional intonation in humans. This warning signal might project directly to the alarm announcing vocalization circuits (in PAG), to swiftly warn the conspecifics of impending danger. Another evidence for the warning role of the dACC is its activation during subjective experience of social rejection [12] and the impaired avoidance learning of harmful stimuli in dACC lesioned animals [13].
The majority of studies results support our claim that the dACC detects when things go bad for us and generates warning signal in both conscious and unconscious monitoring of risky situations and prospects. Dorsal ACC learns and warns about affectively bad choices and consequences (pain, harm, loss, failure) and also about cognitively wrong/failing options, predictions, actions or behaviour. The dACC is causally involved in inhibitory self-control, as it identifies and helps to avoid/stop the bad or suboptimal choices and decisions, which could decrease our survival chances and well-being. dACC is known to be activated by negative affect and cognitive control [14], social and physical pain [15]. Its response is correlated with the frustration felt by participants about their errors in stop-signal task [16]. So pain, bad outcomes and generally anything what went wrong - and we care - are assessed by dACC, leading to inner warning, alarm response, worries and negative affect of distress. Further findings supporting our claim come from dACC lesioned patients [17], which do not increase sympathetic activity toward mental stress. Demanding situations and danger normally activate sympathetic system via dACC, to increase heart and respiration rate. When dACC is in overdrive i.e. by existential chronic worries or by dysfunctionally low dopamine, serotonin or opioid levels, then the repeated warnings if not attenuated by vACC output, cause anxiety and lower baseline serotonin. Because of its warning role toward potentially harmful or suboptimal outcomes, the dACC responds strongly in conflicting, costly, ambiguous or uncertain situations - as they are prone to lead to errors, loss or harm. They contain higher risks of selecting incorrect or inadequate option or wrong cognitive model of the world to act upon – if solving insufficiently defined task or guessing the right response from incomplete information. For the pgACC, active in regret, sorrow and sadness tasks, we predict similar LHb and VS projections as for dACC.

**Anterior insula.**

Multiple studies showed anterior insula activation in pain [18], discomfort, disgust [19], aversion [20], empathy with pain [21], drug craving [22] and in appraisal tasks. The AI receives input from the visual 'What' pathway and from all sensory cortices which bring information about identities or attributes of things, i.e. from taste cortex in the middle insula and interoceptive cortex in posterior insula. The AI primarily identifies noxious or unpleasant stimuli linked to pain, possible contamination, sickness or discomfort. It projects to pre-SMA and premotor cortex to stop response to harmful stimuli. The AI also projects to the cognitive rostral and lateral PFC, VS, amygdala and OFC [24]. It has reciprocal projections with temporal poles which code conceptual information, identities and meanings. Direct efferents of AI may induce gastric motility, nausea toward sickening things or facts.

We propose that the AI primarily detects what is BAD or WRONG - harmful, unpleasant, inferior - in the QUALITY of things (persons, deeds, social conduct) to avoid what makes us feel unwell or sick. The AI reacts with aversion to biologically noxious stimuli which cause pain, contamination, discomfort or bodily harm. It also induces aversion to the bad, low, degraded or corrupted qualities of things including aesthetic, social, moral wrongness. Some such attributes are: odd, strange, misfit, ugly, false, unfair, suspicious, inconsistent, inappropriate, incongruous, corrupted, faulty, sick, disgusting, repulsive or disharmonious. The unpleasant or noxious stimuli – the approach of which might affect our well-being and health, evoke negative affect, discomfort which serves to learn to avoid them. Over-stimulation of the AI possibly lowers the pain threshold, making us feel unwell. Strongly aversive stimuli, rejection or loss of loved ones might evoke even nausea. The magnitude of pain unpleasantness correlates with AI and dACC activation [24]. The phylogenetically oldest, innate function of AI might be to make us avoid noxious food and contact with infected, genetically or disease-distorted bodies, linked to our biological pursuit of well-being and aversion to things which might harm our or our progeny's health and survival chances. This first AI function linked to inborn aversions was extended by social, sexual and cognitive appraisal and judgment.  AI learns to react also to culturally unacceptable behavior and conduct. It detects also qualitative flaws and inconsistencies in information - what is odd, not fitting, false or a lie. This appraisal function is evolutionary linked to the avoidance of what is potentially bad or harmful, depending on its biological, social or subjective meaning, to minimize the exposure. The AI is active in social reasoning tasks, assessing the bad or low quality, aptitude or adequacy of social conduct - targeting what should be avoided or rejected. It is recruited in social cognition tasks when making judgment about outliers, outcasts, injustice (corrupted justice), unfairness or depravity, in moral judgment on what is considered wrong to do to people, in looking down on someone, in rejection. The study [25] where AI activity correlated with rejection of unfair offers, possibly caused by their low quality and by the unfair unacceptable conduct - supports our idea. In decision making and appraisal tasks the AI and dACC often co-activate, exchanging the information about adversities, both capable of response bias toward inhibitory avoidance. While the AI signals bad/low qualities of things to avoid, the dACC warns us when we are not doing well or failing, when getting bad or suboptimal outcomes to our actions, decisions and predictions, relative to the remaining options. So in the context of the NoGo task, the AI identifies the stimuli meant to be avoided, and forwards this information to the dACC, which generates a warning to improve performance.

Diffusion tensor imaging findings (Vadovičová, 2014) showed that AI forms functional processing module with adjacent caudolateral OFC (clOFC). They are often co-activated, strongly interconnected, with shared cortical connectivity, as they probably process similar kind of information – about bad, harmful, aversive, low or wrong qualities of objects, subjects or conduct. The clOFC might do more abstract appraisal and cognitive judgment of what is considered a wrong target, choice, thing to do - in the context of current task, goal, strategy, rule. Thus the clOFC might hold in working memory the information on current distractors and currently inappropriate responses. What to avoid in the current context, to solve a task or reach a goal depends on situation/rules, and it changes with changing circumstances in our physical and social world. So the bad or inferior biological and social qualities are probably first represented in the AI, and the more cognitively assessed and context-dependent wrongness is possibly coded in the clOFC, both exchanging information. Some evidence for this is a joint activation of dACC, clOFC and AI in inhibitory self-control, in NoGo task. This joint role of AI and clOFC in aversive processing is supported by anatomical connectivity study [26] which found that part of clOFC represents a rostral extension of the agranular insular cortex onto the orbital surface.

## AI and dACC input to LHb and D2 loop of ventral striatum form circuit for inhibitory avoidance and self-control.

Most data on properties of the functional circuit, which processes information about adversities and bad outcomes, comes from neuroimaging and psychopharmacological studies. Many fMRI studies show robust co-activation of dACC and AI/clOFC in tasks involving pain, discomfort, harm, loss, error, failure, suboptimal outcomes or threat. What these negative or aversive events have in common is that they are harmful to us, decreasing our survival chances. We learn to avoid them by deselection of the choices linked to the negative consequences. This deselection is done both willfully - by changing our goals and choices in PFC, and implicitly – by probabilistic learning and suppression of bad, wrong or suboptimal choices in the cortico-striato-thalamo-cortical pathway. The third way to cause deselection of bad, painful or punished choices is by inhibition of dopamine and serotonin release (in VTA, SNc, DRN and MRN) via LHb activation, affecting consequently all brain regions with dopamine or serotonin receptors, causing a drop in motivation/drive, hope, cognitive flexibility and well-being.

Affective processing of adversities in the AI and dACC biases/guides value-based choice behavior and learning. Both regions forward information about adversities to the LHb, via which they suppress dopamine and serotonin signaling. This is done directly - by their cortico-habenular projections, and also indirectly - presumably via D2 loop of the VS, which by its GPe projections disinhibits GPi, leading to LHb activation, as GPi is the main source of LHb input. Additional slightly faster prefrontal cortex projections to the subthalamic nucleus stop the wrong response via exciting GPi.

Our DTI study [27] found the predicted cortico-habenular projections from dACC, pgACC, AI and clOFC in humans. Temporal co-activation of these regions in independent component analysis (ICA) of brain functional connectivity (unpublished data) also supported their connectivity, presumably because they form functional adversity processing circuit (APC) selective for inhibitory avoidance learning and self-control. Similar dACC and insular projections to LHb were found in rats [28] in addition to their amygdalar, lateral septal and nucleus accumbens (NAc) projections. A tracing study in rats [29] showed dense cortical projections to MDT, making en-passant and terminal projections to LHb from the AI, ACC, prelimbic and much less from infralimbic cortices. The homologue of rat infralimbic and prelimbic cortex is human vACC and dACC. The prefrontal cortex is known to be strongly interconnected with MDT, forming cognitive and affective loop passing via dorsal and ventral striatum.

Wide evidence for the LHb role in adversity processing, inhibitory avoidance and self-control comes from animal studies. The main input to LHb is from GPi [30]. Other afferents are from lateral hypothalamus, a region involved in nociceptive processing and hunger [31] and [32]. The VTA/SNc and DRN/MRN connections with LHb are reciprocally inhibitory. VTA and DRN suppress LHb [33] and main efferents of LHb inhibit dopaminergic [34] and serotonergic neurons [35], directly and via rostromedial tegmental nucleus projections RMTg [36]. The LHb projects also to supramammillary nucleus (SUM) [37]. Macaques electrophysiology [38] gave evidence of LHb activation by aversive outcomes, punishment and reward omission. These neurons were most excited by the most negative of the available outcomes, firing when the outcomes were worse (less rewarding, more punishing) than predicted, thus inversely to VTA/SNc neurons. Because the LHb inhibits dopamine and serotonin signaling, it affects our choice behaviour, learning, motivation, moods, well-being and self-control. The proposed APC gets activated when we are not doing well or things go wrong. It identifies the adversities via AI/clOFC and dACC and forwards this information to LHb, to inhibit dopamine and serotonin release. It learn to avoid adversities by stopping choosing/doing things that hurt, lead to negative consequences or lower survival chances.

Many human studies [39]-[41] demonstrated stimulating effect of dopamine (DA) on D1 loop of the basal ganglia and on learning to repeat good/rewarded choices, versus its inhibitory effect on D2 loop and learning to avoid maladaptive choices. They found that high DA levels caused D2 loop inhibition and disruption of inhibitory avoidance learning, while low DA potentiated its strength. The principal striatal neurons express D1 or D2 receptors, and project to the GPi via direct D1 (inhibitory) versus indirect D2 (excitatory) loop. They have opposing effects on the MDT: the D2 loop inhibits and the D1 loop disinhibits it [42].

By linking previous facts together we concluded that the D1 and D2 loops of ventral striatum differ in their cortical input, depending on the type of affective information they receive, evolutionary linked to either approach or avoidance. The expectation of something good, valuable, rewarding biases and motivates us to seek, engage with, approach, go for and choose this valuable choice. While the expectation of bad things or consequences – such as pain, loss, harm, punishment or failure demotivates and stops/inhibits us. Because we are evolutionary biased to seek, move for and expend effort toward the valuable – survival promoting stuff and to avoid, stop choosing/moving for that what tends to harm us. The striatum uses probabilistic implicit learning, to strengthen the glutamatergic synapses depending on good versus bad outcomes and expectations, signaled by DA levels. So the motivational bias in the VS depends on both the input strength and the strength (weights) of the synapses linked to each choice/option, and normally takes more trials to update. Thus a strong cortical input plus strong DA signal at D1 receptors might pass even through weaker synapse. So your preferred choice can still win in VS even after getting more negative than positive feedbacks (that strengthen D2 synapses) if you strongly value or believe in it (i.e. when in love). And strong VS synapses are insufficient if the cortical input is weak, meaning if the expectation of reward (i.e. in depression) or punishment (i.e. when drunk) is weak.

Our reasoning is that the cognitive part of the PFC, which is flexible to integrate/combine and hold in working memory any kind of information, projects equally to both D1 and D2 loops of the cognitive DS in the caudate head. By summing the cumulated evidence given by ''right'' versus ''wrong'' feedbacks that strengthen D1 versus D2 loop, the correctness versus wrongness of options, predictions or instrumental rules is learned by the DS. Those predictions, guesses or ideas about relevant causal relations, task rules (i.e. IF...THEN rules), strategies, or instructions (manuals) on 'HOW TO DO' things, which get most consistently confirmed or approved by incoming evidence (experiences), gradually win over the less supported or more disproved options in DS. The positive feedback caused by correct or right outcomes acts through the DA peaks it induces in the SNc. These DA peaks then enforce the synaptic strength of the relevant pathway/option in the D1 loop, while weakening the strength of the parallel and competing D2 pathway. Conversely, negative feedback causes a drop in the DA: enforcing the D2 loop of the DS and the probabilistic learning of what is the wrong way of doing things or incorrect rule/strategy. So in the head of caudate are counted all 'Yes' and 'No' feedbacks, to weigh all the evidence FOR and AGAINST each tested proposition, prediction or instrumental rule, by Bayesian, probabilistic learning. By counting trials success and errors, it teaches us what is right or wrong way, directing us implicitly, i.e. when learning grammar rules.

Cortical input to the VS differs from the lateral head of caudate (DS), because it provides information on the affective meanings or values of choices - expected to be good, bad or combined. The general role of basal ganglia is to give directional guidance for actions and choice selection (i.e. go right vs. left, go for it vs. avoid, right vs. wrong way to do), depending on weighted sum of our good versus bad experiences with each option, thus on the DA signal.

The input from affective cortical regions to VS, about good, liked versus bad, disliked choices, already contains the directional information: a BIAS for approach versus avoidance, linked to their value for us. Because we are evolutionary biased to seek, approach, move for and choose the good - valuable, rewarding, interesting – survival promoting choices, and to avoid/stop going for the bad or harmful choices - linked to pain and lower well-being or survival chances.

So we propose that the cortical input to the D1 versus D2 loop of VS depends on the affective value of the forwarded information: the expectation of good, valuable things feeds preferentially to the D1 loop, while expectation of bad things (i.e. punishment) feeds to the D2 loop. Thus the mOFC and vACC input, about good, positively valued choices, is preferentially forwarded to the appetitive D1 loop, to guide us to choose, seek, GO, move and work FOR the valuable options. While the information from the AI, clOFC, pgACC and dACC about bad, harmful, suboptimal, negatively valued choices feeds predominantly to the aversive D2 loop and directs us to avoid and stop moving or making an effort for them.

This affective value-based dissociation of cortico-ventrostriatal input can be tested by optogenetic neurostimulation studies. There is also ample indirect functional evidence from choice behaviour studies, dopaminergic plus serotonergic manipulations and findings on addiction, depression, bipolar disorder, anxiety or Parkinsonism, that fits the proposed connectivity and functional properties of the APC. Indirect evidence is that the cortical regions that learn from bad versus good choices and consequences – activated when things go wrong versus well – cause activation versus deactivation of the ventral GPi and LHb, followed by suppression versus induction of dopamine and serotonin signaling in the human brain. Proposed APC properties were supported by our DTI (27) and ICA fcMRI findings (unpublished data), by the GPi response to expected punishments in macaques [30], by decreased versus increased compulsive drug-seeking after optogenetic activation versus inhibition of the dACC homologue in rats, as well as by lower dACC excitability in the cocaine-addicted vs. non-addicted rats [43]. These findings support the idea that dACC and AI/clOFC are sources of the direct and indirect input to the LHb, with which they form a functional circuit, which learns to de-select or avoid what is bad or unpleasant for us. So it learns to inhibit bad/suboptimal choices and actions that hurt, are punished or lead to painful outcome. We suggest that affective cortical evaluation in the vACC, mOFC, AI/clOFC, dACC and the strength of their input to VS linked to each choice depends on our subjective values, preferences, priorities, expectations and interpretations of what is good or bad for us. Those choices or options which bring a combination of good and bad outcomes, will feed to both the D1 and D2 loop of VS. For example food mostly activates the D1 loop, but in satiety, dieting or in anorexia it also activates the D2 loop, when the normally rewarding food choice also becomes something "to avoid". The role of mOFC in coding the reward value of stimuli - their relative reward magnitudes was proved by many studies. mOFC responds to rewarding or pleasant stimuli and wins, while the lateral OFC responds to aversive options, punishment or loss [44]. Another support for the proposed dACC and AI input to the LHb comes from their co-activation during negative feedback in human fMRI study [45].

Directionality coding in basal ganglia uses probabilistic learning to select the best - most rewarded options and ways of doing, while avoiding loss and harm. Good, rewarding, interesting options bias the choice selection in VS toward themselves by potentiating the D1 loop by DA - DA being induced by positive expectation and good feedback. In this way, good choices/chances make us engaged, interested, motivated, "hooked on" to want, seek, move, make effort and do things to get them. Bad, painful, harmful or boring choices (loss of time, unpleasant to do) are linked to inhibitory avoidance ("Don't do it, Stop"), to drop in motivation, even learned helplessness, caused by D2 loop potentiation in VS by lack of DA after negative outcomes or negative expectations. As DA is known to weaken the D2 loop output, a drop in DA release, caused by the APC activation and by negative feedback, leads to strengthening of the inhibitory avoidance linked to suboptimal choices, decreasing their preferences or inclination to choose them over time. Thus, DA increases drive and inclination for rewarding options in the motivational D1 loop of VS: their wanting, seeking, choosing to GO FOR them. DA also potentiates the D1 loop of cognitive DS that gathers evidence on what is probabilistically more RIGHT, correct option, WAY TO DO things or an optimal course of actions to reach the task goal. Meanwhile, the D2 loop of DS, potentiated by negative feedbacks, keeps evidence on failed/suboptimal ways of doing things and on decisions/predictions that went or were proved wrong.

Similar dissociation is found in the effect of affective information (and their evaluation) on the VTA versus LHb function. While the expectation of good, valuable – interesting, novel or rewarding things and outcomes formed by mOFC directly stimulates DA release, the expectation of adversities and bad outcomes formed by AI/clOFC and dACC inhibits VTA/ DA and DRN/5-HT release via projections to LHb which activates the RMTg. Inhibition of dopaminergic VTA/SNc output by LHb [46] and the sustained increase in the striatal DA after LHb lesions [47] are well evidenced.

We propose that the activation of the D1 loop of VS by mOFC, vACC, hippocampus and amygdala input inhibits the GPi and disinhibits MDT, what enforces the representation of valuable choices in the PFC. Reward based GPi inhibition decreases   LHb output so disinhibits VTA/DRN, as GPi is the main LHb afferent [48]. The activation of the D2 loop, by dACC and AI projections, disinhibits GPi, leading to MDT inhibition and LHb activation causing DA/5-HT suppression. So it potentiates the passive avoidance learning in the D2 loop. The network, formed by linking affectively negative input about adversities from the dACC and AI/clOFC to D2 loop of VS that disinhibits GPi and consequently activates LHb, can also be called the inhibitory avoidance circuit. As it biases us to avoid or stop choosing the bad or suboptimal options.

Because the LHb reciprocally inhibits the VTA/SNc, the net effect of the APC activation is opposite to the RPC activation. So both the adversity and reward related circuit act in parallel and in competition to influence learning and bias motivation, decision making and choice behavior toward avoidance versus approach – STOP versus GO response, 'Do not do it!' versus 'Do it, Go for it'. In addition we claim that when things are going well, alright, when we gain, reach, achieve what we wanted (safety or rewards) the vACC projections to DRN induce serotonin release. When things go wrong, with bad/suboptimal outcome, the dACC and AI/clOFC inhibit the DRN and VTA, via LHb projections, inducing discomfort. While the vACC identifies good outcomes and safety, making us feel secure, content, the dACC appraises risks and dangers - inducing warning, worry and alarm state. Our conclusion is that the RPC and APC are competing in the VS for control of the GPi, LHb, VTA, DRN and choice selection, guiding our affective, value-based decision making.

## The APC versus RPC: their competition and interactions with dopamine and serotonin signaling.

Multiple findings support the causal role of the APC in learning what is bad/ harmful to us and reacting to it by inhibitory avoidance and inhibitory self-control. The proposed competition between RPC and APC - by their reciprocally inhibitory connections and opposing effects on dopaminergic and serotonergic control, fits with wide evidence from neuroimaging, psychopharmacological and anatomical studies. Also our idea of mostly activating effects of dopamine on RPC, and its mostly inhibiting effects on APC, serving to bias/direct our choice behavior, is supported by evidence.

Because the LHb, the target of the AI and dACC, inhibits VTA/SNc and DRN – and is reciprocally inhibited by them, the net effect of the APC activation is antagonistic to the RPC activation. In this way, by the opposite effect of dopamine and serotonin on the activation of each circuit – are the RPC and APC competing to bias our decision making, motivation and well-being. They guide us toward selection, approach or toward de-selection, inhibition, avoidance of the current choice, option or goal. Thus to 'Go, Move for it, Do it' versus 'Stop, Avoid, Do not do it' response. Dopamine release acts to increase the inclination for the rewarding choices and goals in the motivational D1 loop of VS - their wanting and seeking. DA also increases selection of the more frequently right/correct options or courses of actions in the D1 loop of cognitive DS, and potentiates implicit learning of the appropriate stimulus-response mapping in the sensorimotor loop of putamen via D1 loop.

There is an additional network-level competition in the brain - caused by the interaction of dopamine levels with the output strength of these two opposing affective circuits. Dopamine in basal ganglia enforces the strength of the D1 loop but attenuates the D2 loop. Similarly, we propose predominantly excitatory versus inhibitory effect of DA on the cortical regions of RPC versus APC. This network effect is likely achieved by a higher ratio of the excitatory D1 to the inhibitory D2 receptors on the GABAergic neurons of APC, combined with their lower ratio on the pyramidal neurons. These ratios are probably inverted in RPC. So DA attenuates the appraisal and predictions in AI/clOFC and dACC about what is bad for us, leading to lower LHb recruitment. This is done by prevalence of the inhibitory D2 receptors on their pyramidal or the excitatory D1 receptors on their GABAergic cells. The contrary is predicted for the cortical regions of the RPC: mOFC and vACC, where DA likely strengthens the output and learning of what is good for us, leading to stronger mOFC to VTA/SNc input. Except the VTA/SNc - with its inhibitory autoreceptors to implement negative feedback control, the remaining RPC regions are mostly enforced by higher DA levels and attenuated by DA absence. Our observation, based on ample literature, is not only that the activation of APC leads to suppression of dopamine and serotonin release, but also that the APC output is generally attenuated by dopamine, serotonin and opioids.

Important evidence for the proposed interaction of the affective value-based processing with dopamine and serotonin signaling follows. High DA levels are known to decrease inhibitory self-control and cautiousness. Our causal explanation for this is that DA attenuates dACC output. Because dACC is involved in monitoring and WARNING us about possible dangers (Beware!), risks and negative consequences (to signal the inadequate course of actions), the suppressing effect of DA on dACC (via D2 receptors on glutamatergic or D1 receptors on GABAergic neurons) leads to lower precaution, premeditation and self-control, less worries and less thinking before doing. This underestimation of possible risks and bad consequences – caused by attenuation of dACC output, diminishes the urgency of inner warning signal during appraisal and decision-making, and lowers the inhibitory self-control during performance. The impulsive or suboptimal choices/behaviours (those leading to bad or suboptimal outcomes) might then win more easily over weakend self-control. This commonly happens when DA in brain is high - caused by alcohol intake, dopaminergic drugs, high reward expectation or by DA imbalance in several mental disorders. There are many documented examples of lost precaution or underestimation of risks after alcohol intake – drunk driving, random violence, wrong or stupid decisions, or in manic phase of bipolar disorder - high spending. The reduction in dACC output and its warning signal, caused by high DA, weakens the D2 loop output, and thus relatively enforces the D1 loop output in VS – leading to impulsive choices. Acute alcohol intake stimulates DA release in VS and impairs executive functions of PFC [49]. Even moderate alcohol consumption reduces activity in dACC in response to errors of performance, what impairs the ability to adjust behaviour after such errors [50]. We claim that the known effect of D2 antagonists in reducing the positive symptoms of schizophrenia, is partially caused by their disinhibition of dACC – which leads to increased warning function, precaution, self-control and worrying, while the disinhibition of D2 loop of VS shell increases inhibitory avoidance, reducing the impulsivity and drive.
Strong evidence for the proposed affective circuit properties is a finding [51] that dopamine reward signal from VTA suppresses the error-related negativity ERN in humans, decreasing its amplitude. Our interpretation of this effect is that the ERN is caused by the warning signal generated by dACC – toward potential danger or risks. And that high DA levels attenuate dACC output, leading to lower ERN amplitude, less precaution and self-control, plus weaker cognitive adjustment – linked to lower dACC feedback to lateral PFC. Further evidence comes from a study [52] where positive symptoms schizophrenics failed to activate LHb in response to errors and negative feedbacks in demanding cognitive task. Authors showed that this dysfunction hinders patients' ability to learn from negative feedbacks, what is a known cognitive deficit in schizophrenia. We claim that one cause of the LHb under-activation in schizophrenia is the suppression of its cortical input (from dACC, AI and clOFC) by high DA levels. High DA weakens the warning signal and the weight of bad consequences taken into account during decision making. Humans might then fail to adjust behavior to prevent loss or failure. Depression is linked with larger ERN during error processing and with better avoidance learning [53]. We interpret this finding as over-stimulation of the APC after its disinhibition by insufficient dopamine and serotonin levels in the depressed brain, causing stronger inhibitory avoidance.

We propose that serotonin in brain generally attenuates adversity processing, negative affect, discomfort, deprivation, depression, anxiety, worries - by suppressing dACC, AI/clOFC and their input to LHb. Analogous to its calming effects on pain pathway, serotonin placates also threat response, anger, fear and aggression by its inhibitory effects on central nucleus of amygdala (CeA) and possibly bed nucleus of the stria terminalis (BNST). In support for proposed circuit properties is the high density of the inhibitory serotonin receptor 5-HT(1A) found in dACC [54], the fact that 5-HT(1A) agonists ameliorate anxiety disorders [55] and a negative association between the activation of dACC and local 5-HT(1A) mediated inhibition [56]. Knocking-out this receptor in mice led to anxiety-like behaviour [57-59] and its lack during development [60] induced anxiety, resembling the effects of social deprivation on human babies. There is wide evidence for appeasing effects of 5-HT on violent behavior, which we explain by its attenuation of CeA. The aggressive behaviour correlates with low 5-HT levels in humans [61] and is decreased by the agonists of several 5-HT receptors in rodents [62]. Similar inhibitory effect on dACC has also mu-opioid, an opioid with anxiolytic, calming effect on worries, discomfort and pain. High mu-opioid receptor density was found in dACC, perigenual ACC, all insula and LHb [63], [64]. Thus besides the subcortical pain circuit, it inhibits also APC. As mu-opioid disinhibits VTA dopamine and DRN serotonin neurons through inhibition of GABAergic RMTg terminals [65], it enforces the strength and output of RPC and lowers that of APC.

Serotonin has different effects on different RPC regions. We predict inhibitory effect of serotonin on DA release in ventral striatum, leading to attenuation of D1 loop output. We predict excitatory effect of 5-HT on vACC region, via pyramidal 5-HT(2A), 5-HT(2C) or other receptors. These receptors are reduced in major depression disorder [66]. Their activation lowers stress response, anxiety and depression. But main reason for our prediction is the ample evidence that both vACC recruitment and serotonin signaling are linked to well-being: feeling alright, at ease, safe, fulfilled, in peace - leading to drop in wanting, deprivation, drive, to slowing down and rest. We propose that vACC identifies when we reach/gain safety, comfort, valuable resources, biological and social rewards, affection, care, acceptance or good feedbacks, then promotes satisfaction, confidence, socializing. This appraisal in vACC that things are going well for us induces serotonin release via its projections to DRN/MRN. These projections were proved [67]. vACC might activate DRN also via BNST or BNST/RMTg. Serotonin strengthens the D1 loop of cognitive DS, biasing the decision making toward the cognitively optimal or right options (bigger delayed rewards, better long-term goals). The reasonable choices are competing for selection (in MDT or PFC) with the pleasant immediate rewards coded by D1 loop of VS shell - linked to impulsivity, wanting, drive (drive and motivation is linked also to medial DS). Sufficient serotonin levels promote optimal deliberate decisions by attenuating the affective D1 loop output and impulsivity by lowering DA signal in NAc. Deprivations and despair bias decision for impulsive choices, to get reward to survive ''right now''. Serotonin enforces cognitive selection of the optimal course of actions/behavior for current problem or goal by increasing DA signal in cognitive DS loop (by suppressing GABA input from SNr to SNc). Serotonin might also attenuate mOFC, thus lowering the relative value of things we reached - got satiated with, to reduce wanting and seeking when we had enough, to relax rest and recover instead. So 5-HT changes motivation and priorities by signaling when we reached the comfort zone of homeostasis – being well, fine, full, non-deprived.

Majority of findings in literature are in accordance with proposed network competition in affective processing of good versus bad choices and outcomes. For example noxious heat deactivates vACC and mOFC [68] and activates AI, dACC and LHb. The social support, which includes social gain, care, affection, acceptance and group inclusion, causes a decrease in AI and dACC activation [69], what we explain by the activation of competing vACC region by care and affection. vACC inhibits dACC, increases comfort and well-being, plus causes serotonin rise, which then attenuates dACC, AI, LHb and amygdala output. Chronic overstimulation of the APC, caused by too many adversities and negative feedback (things going wrongly, harm, loss, punishment, pain, rejection) combined with the lack of positive feedback to balance it (or an underactive RPC or weak 5-HT signal) can lead to excessive suppression of DA/5-HT release (via LHb) and to a long-term shift in their basal levels. This causes a drop in motivation, hope and effort (by lack of DA in D1 loop of VS), lowers the well-being and pain threshold (weakening the vACC and strengthening pain circuit by lack of 5-HT in brain) and increases discomfort (by disinhibiting AI), learned helplessness (by too strong D2 loop of VS), worries and anxiety (chronic activation of dACC and CeA).

Those neural dysfunctions which lower dopaminergic and serotonergic signaling (receptor-deficiencies, too efficient degradation enzymes, low synthesis) enforce the output of APC, relatively weakening or outweighing the output of RPC, so the affected persons get more prone to behavioral inhibitions, worries, feeling down/low, pain, stress, depression and anxiety. This idea fits with findings [70] of abnormally reduced DRN region in depressed, nonalcoholic, suicide victims. Thus, low 5-HT signal leads to relatively higher dACC and lower vACC output, decreasing well-being, satisfaction, feeling alright - at ease, and increasing discomfort.  A majority of findings supports our idea on affective networks competition. One group [71] found higher dialysate 5-HT concentrations in the NAc, DRN and VTA, and increased DA levels in NAc and DRN after cocaine challenge of chronically cocaine-treated animals compared with controls. This shows that chronically high DA levels enforce the output of RPC - by relatively suppressing the APC, leading to lower LHb recruitment and thus desinhibition of dopaminergic and serotonergic system. Another group [72] found opposing effect of simultaneous threat and reward presentation on the dACC and AI activity.  Pharmacological evidence shows abnormally decreased 5HT1A receptor function in mood disorders [73]. This can cause disinhibition of worries, anxiety, discomfort and distress produced by dACC, AI, LHb, CeA and paraventricular thalamic nucleus (PVT). The antidepressant effect of optogenetic activation of the medial PFC, which includes both mOFC and vACC region was demonstrated [74]. These cortical regions are main sources of VTA and DRN activation. In addition, a drop in serotonin levels, elicited by a reduction in dietary tryptophan, increases regional cerebral blood flow in LHb that correlates with depressed mood ratings plus anti-correlates with plasma tryptophan levels [75].

Enhanced LHb metabolism and reduced brain serotonin levels have been observed in several animal models of depressed behavior. The first is suppressed by antidepressant drugs [76], the second by LHb lesions [77]. Administration of serotonin reuptake inhibitors induces increase in serotonin levels and attenuates negative introspective processing in subjects at risk for depression [78], showed to be driven by decreased responses to the negative self-referred words in dACC after citalopram compared with placebo. Another evidence for the LHb being an effector region for the APC is that LHb activation increases in stress [32] and by nociceptive stimulation [79], [80] same as the AI and dACC - its input regions. Lesions or morphine microinjections in the habenula decrease animal's sensitivity to pain [81] and their anxiety [82]. We suggest that weakening of LHb disinhibits DA and 5-HT system, shifting the balance from discomfort to well-being, by enforcing the RPC output at the expense of APC and pain pathway. Dopamine system disinhibition increases impulsivity and drive and weakens inhibitory avoidance. Habenula-lesioned rats show a marked increase in premature responding mediated by increased activity of DA neurons [83]. This supports our affective circuit model – and the competition between the 'Go for it' motivational bias in impulsivity promoting D1 loop of NAc shell, and the inhibitory avoidance bias in its D2 loop. Several mental dysfunctions are caused by disturbances in DA or 5-HT signaling, affecting learning, decisions, goal-directed behavior, personality, moods, motivation and well-being. The proposed neural mechanisms of reward and adversity processing, which guide choice behavior depending on good and bad outcomes, led to our following model on dopamine and serotonin signaling and value-based decision making in brain:

### Dopamine signaling.

Dopamine is signaling the expected value and meaning, guiding us to go for and choose them. It signals, that there is something valuable – good, rewarding or novel for which to move, go, get and make effort. It also signals meanings in our world and life – what is relevant, right, valid, useful, interesting. DA increases our drive, hope, wanting, impulsivity (a bias to choose immediate rewards in spite of bad consequences), engagement with valuable choices, approach, preference and addiction by enhancing the motivational D1 loop of VS. DA signaling makes us motivated (hooked on) and directed (inclined, biased) toward choosing, seeking, moving for, working for and learning about the valuable - survival and well-being promoting stuff and choices, those which are reinforced by DA. So biologically, socially, personally or cognitively valuable things, persons, deeds, events or information are those which increase our survival chances or well-being, when gained or reached. The things needed for survival are usually liked, appraised as good and feel rewarding, pleasant, exciting, joyful (food, relationships, beauty, novelty) or interesting (novelty, variability, information, knowledge).

The relative magnitudes of reward values are identified, learned and predicted by mOFC, which projects its reward expectation signal to VTA. Sensory novelty and variance shows richness of the environmental resources so is intrinsically coded as valuable, by increasing the mOFC activation. The informational novelty is also valuable, because it motivates us to seek, guess, figure out new causalities in the changing world, leading to useful findings of what is going on. The acquired knowledge might help to obtain new resources, gain safety or avoid harm thus improve our survival chances. This informational novelty is probably detected by lateral BA 10, which induces DA release in medial SNc, provoking novelty seeking and curiosity, thus drive and motivated behavior via D1 loop of VS. The relevant, meaningful information - in the context of current goal/task (useful pieces of spatial, temporal, objects or meanings related information) are coded by PFC and induce DA in SNc, to signal the informational value and to enforce the representation of the most relevant information in working memory.

Dopamine, released in DS after detection of the valuable or correct options by PFC, increases the probability of selecting the right options again, by enforcing synaptic efficiency of corresponding D1 loops and by strengthening their cortical representations. So DA strengthens working memory in cognitive PFC for meaningful, valuable information, relevant for current goals. DA enforces also the relational (things bound into temporal, spatial and relational context, cognitive maps) memory in hippocampal CA 1, for rewarding, interesting, relevant or novel features/events.  Also subcortical projections induce DA in SNc. PPT efferents increase alertness, while input from superior colliculus evokes DA response to salient or abrupt visual events [84]. So DA signaling serves to learn about and direct us to valuable and meaningful – good, right or interesting choices in ever changing biological and social world, while its lack potentiates behavioral inhibition. DA signaling leads to flexible behavior by guiding us to choose/prefer the best options - good, right or interesting choices, goals, decisions, predictions, behavior while avoiding bad, wrong or suboptimal choices. The value of some things is inborn the value of others is learned and updated by experience, by our interactions with world and people.

Knowledge seeking, curiosity, a desire to find out what is going on, and creativity, playfulness is DA-related and probably affected in autism. The ability to ask questions, to explore is the brain's tool for model free learning, as the unknown makes us seek 'Why' - causes, sense, meanings. It directs us to think, guess, predict, test and learn whatever is interesting, useful, relevant in whichever situation or environment - making us versatile specie, able to solve new challenges and adapt. As DA release in SNc is caused by informative, goal-relevant, meaningful (what matters to us), interesting or novel information (appraised by cognitive PFC) it is important for flexible learning/use of the currently valid, correct rules, predictions or ways of doing things. Random release of DA in psychosis state leads to potentiating/learning irrelevant information as if relevant and meaningful, thus seeing meanings in unrelated facts or random events and validity in invalid ideas, affecting judgment. So learning and behavior in psychotic states is disconnected from reality and immersed in artificial reality caused by dysregulated DA signaling (which confers meaning) and by attenuated warning signal (in dACC) towards bad or suboptimal decisions.

Addictive drugs increase DA levels in VS few times more than natural rewards do, so the drug gains the priority in front of other attractive rewards, choices or goals. The value of the remaining choices then relatively drops and addicted person loses interest to work for, seek and strive for other rewards or goals than the strongest one - drug. So the drug gets encoded on the top of the personal values and priorities scale in the mOFC. Drug-linked path in D1 loop of VS wins over other choices and goals because of its strong mOFC input plus strong synaptic weights. As the competition in VS biases us to go for, seek the winning choice, it causes dysfunctional choice behavior in drug addicts, added by drug induced rise in DA which reduces self-control and inhibitory avoidance in APC. In accordance with this, cocaine induced 'high' evokes VS activation [85]. And blockade of D2 receptors in VS by neuroleptics was found to cause anhedonia, apathy and psychomotor slowing [86]. Thus D2 loop disinhibition causes inhibitory avoidance and overstimulation of LHb, causing drop in DA and motivation.

We predict that oxytocin stimulates the motivational D1 loop of VS core by presynaptic increase of DA release - to make us interested, engaged and attached/bonded to specific persons (kids, friends) to prefer want and seek them. Oxytocin probably also stimulates vACC. Diminished oxytocin signaling was found in autism. We propose that insufficient DA signaling in autistic persons makes them worse in value-based learning, less driven by rewards, novelty and meanings, with low curiosity or novelty seeking, thus failing to find out causalities, make predictions, adapt – making their world confusing, unpredictable.

**Serotonin signaling.**

In opposite to DA signaling which makes us want, hope, seek, strive and move for the valuable stuff, serotonin signaling suppresses drive and wanting, slow us down, calms anxiety, fear and stress, inducing contentment. We propose that serotonin in brain is released when we gain, win or achieve biological and social rewards, resources (food, safety, comfort, care, affection, acceptance, smile, success), to make us feel well, safe, satisfied, comfortable, confident, fine, at ease, peaceful, full - so ready for rest, digestion and body recovery. We predict that the vACC/BA 25 detects and signals when things are going well or alright for us - when we gain safety, valuable resources, rewards, good feedback, reach goals, do something well or improve our prospects. vACC induces well-being and satisfaction by inhibiting stress regions and activating serotonergic cells in DRN, causing suppression of pain pathway, threat response in central amygdala and worries in dACC. Postsynaptic 5-HT(2A) receptors are target of several antipsychotics, anxiolytics, antidepressants and hallucinogens, for example MDMA (3,4-methylenedioxy-N-methylamphetamine) or "ecstasy". The MDMA is also a potent releaser and re-uptake inhibitor of presynaptic serotonin [87]. These effects (except the hallucinogenic effect in retrosplenial cortex) are probably caused by strong vACC activation, which leads to further 5-HT release, if vACC wins over dACC output strength. Shifting the balance between strong dACC output and its warning role, toward strong vACC output – and its calming role, attenuates worries, stress, fear and anxiety. Thus optimal brain serotonin levels signal that we are non-deprived while its lack signals deprivation state.

In support of this, a positive correlation was observed between tryptophan (source of serotonin) levels and response to reward outcome value in human ventromedial PFC [88]. In this region with dense 5-HT(2A) receptors was found 5-HT efflux in instrumental reward task in rats, and its activation by either 5-HT agonist or AMPA elevated 5-HT levels [89].

We propose that serotonin's function is to keep body homeostasis in healthy, life-compatible limits by controlling our intake, promoting rest states, well-being and drop in wanting. It signals when we reach the 'comfort zone' - satiety, satisfaction, security, comfort, care, affection, good feedbacks/outcomes, acceptance. Also when in pleasantly warm/sunny, hospitable, pretty, rich or resourceful environment, blooming nature, nice home, lively town or company (contrary to the destroyed barren boring inhospitable places devoid of resources). Too low serotonin levels signal that body and mind are in a state of deprivation, discomfort or despair, caused by pain, hunger, loss (of loved ones, health, work, safety, abilities, control or future prospects). Depletion of serotonin lowers the pain threshold and there is a mutually enforcing effect between serotonin and opioids as both attenuate pain perception and lift up moods. Besides well-being, serotonin regulates also our choice behavior. When our needs are fulfilled and goals achieved - when satiated, safe, nurtured, cared for, liked, serotonin levels rise and attenuate via inhibitory receptors the strength of the D1 loop of NAc shell, lowering impulsivity, drive, desire, wanting and moving for more rewards. This might be mediated by 5-HT(1B) receptors, as their stimulation was shown to attenuate the incentive motivation for cocaine [90]. It was shown [91] that deletion of these receptors leads to an increased locomotion and sensitization of mice to cocaine self-administration, caused by a reduced inhibition of dopaminergic transmission at the VTA and NAc. VS activation is enhanced by serotonin deficit and correlates with steeper discounted reward choices, while DS activation correlates with reward expectations at longer delays and is stronger at high serotonin levels [92]. Animal studies found serotonergic projections to both the VS and DS [93], [75]. We suggest that natural drive to chase rewards and resources is strongest in the deprived state when lacking serotonin in the brain. After gaining what we wanted, getting fed or full, the induced serotonin reduces wanting, impulsivity, drive, effort - leading to slowing down, rest and body recovery. This has a biological sense as people deprived of resources, desperate – by lack of food, social bonds, good outcomes of own actions - need to act urgently to survive 'right now' in spite of pain, punishment, risk or inconveniences involved (disinhibited D1 loop wins even over strongly active D2 loop in VS). So non-deprived persons with enough serotonin in brain can wait for delayed rewards, but its depletion heightens the impulsivity by disinhibition of D1 loop in NAc, to go for valuable things immediately.  Serotonin in brain regulates also food intake, by decreasing appetite in arcuate nucleus of hypothalamus. Increased food consumption is probably linked to lower basal serotonin levels [94] while inhibition of 5-HT re-uptake leads to weight loss in obese individuals [95].

The reports of DRN activation in some pain studies might be caused by the termination of aversive events and the occurrence of safety or relief after the discomfort stopped. We suggest that just after the pain cessation, the vACC appraises that we are doing well now - being safe, then signals this safety and relief information to the DRN to induce serotonin. Another serotonin role is to calm down anger and aggression by inhibitory effect on threat and fear expression in the CeA, BNST and panic-linked part of periaqueductal grey (PAG). The rat homologue of BA 25 (posterior part of vACC) was found necessary for fear extinction learning [96], so for learning about safe options. The fear extinction learning pathway inhibits the CeA fear related neurons by projections from the vACC to the GABAergic intercalated neurons [97]. The inhibition of the DRN increases aggression irrespective of alcohol intake [98]. So the calming serotonergic effect on violence might get selected for by breeding animals with 'domesticated trait' – with relatively stronger vACC signal and higher serotonin levels, combined with lower dACC and CeA output. One support for this claim comes from deficient vACC activation in aggressive individuals, coupled with exaggerated amygdala response to stressful situations [99]. So the individual differences in the vACC output strength, basal serotonin levels or receptor efficiency are possibly linked to personality trait - a stronger activation causing confidence, feeling at ease, relaxed, secure, in control while a weaker one causing stressed, fearful, anxious or impulsive personality traits. Early social isolation stress reduces the vACC output in mature primates, possibly by modification of histones or transcription factors or by decreased BDNF linked growth. Increased aggressiveness, which exaggerated after early social adversity, was observed in the male macaques with genetically lower serotonin signal [100]. The evidence from literature supports serotonin's capability to lower sensitivity to aversive events and stress. We explain this calming effect by 5-HT inhibition of dACC, AI, clOFC, CeA and LHb and by potentiation of vACC. Both effects are anxiolytic and antidepressant, lowering worries and discomfort, leading to feeling safe, at ease, alright, well.

Optimal brain serotonin levels increase cognitive flexibility by making us feel non-deprived, less impulsive, balanced thus more free/flexible to choose the most appropriate, optimal options. Even delayed or long-term goals, as we are not in urgent need to get resources instantly – to survive till the next day. In support of our claim are findings that acute and chronic reductions in central and prefrontal serotonin availability increase behavioral rigidity and reduce the inhibitory functions [101], [102]. Serotonin influences also social and sexual behavior: high brain levels decrease drive, making us lazy, fed up or disinterested, but optimal levels help to feel at ease, fine, in control thus more outgoing to socialize and more successful in social interactions, in winning friends by self-confident, relaxed, trust and respect evoking appeal. It is also easier to affiliate and cooperate when safe, nurtured, comfortable (with own look, abilities), liked than when deprived, scared, under pressure, worried, down. This is supported by increased affiliative behaviour after treatment with selective serotonin reuptake inhibitor [103] and by findings that the hypothalamic 5-HT(2A) receptor activation causes increases in oxytocin, the hormone involved in bonding [104] and trust. Being in control of your life: doing well and getting good feedback stimulates vACC, increasing resilience to stress, as vACC and subsequent serotonin attenuate dACC and amygdala (5-HT might inhibits also PVT).

**Active avoidance loop via amygdala, potentiated by DA.**

The amygdala is well known for its reactivity to biological and social threat, fight or flight response, self-defense, anger, aggression and stress. Its role in conditioned fear learning and aversive emotional reactivity was proofed [105]. We claim that in addition to activating inhibitory avoidance (by their inputs to APC) both AI/clOFC and dACC trigger also the active avoidance, by their inputs to amygdala about appraised threats, risks and potential dangers. The amygdala projects to D1 loop of VS, to BNST and via BNST to PAG, VTA and stress related hypothalamic paraventricular nucleus (PVN). This amygdalar circuit for active avoidance reacts to threat and distress by anger, fear or panic. Contrary to the inhibitory avoidance circuit, we claim that the active avoidance circuit is stimulated by DA and dopaminergic drugs, via excitatory receptors. The BNST has both excitatory and inhibitory projections to the VTA. BNST potentiates DA release also via corticotropine releasing factor. So DA enforces the active-reactive avoidance, anger and defense of self, close ones, territory or vital resources via fight or flight. If people are repeatedly deprived of basic needs or rights and things are going too bad for them, they might lose hope and feel down via suppression of the VTA by LHb over-stimulation. With the insufficient DA levels to excite their amygdala, it gets under-activated, what causes a drop in the active avoidance and diminishes protest or defiance to fight against oppression or causes of their bad conditions. Thus the lack of DA would not only enforce the D2 loop and increase passive avoidance and learned helplessness, but also suppress the self-defense behavior via amygdala, common in depression. On the contrary, high DA levels i.e. under alcohol influence, may lead to violent or irresponsible behaviour, as they potentiate angry or furious response via amygdala, impulsivity via D1 loop of VS, plus attenuate cautiousness via inhibition of dACC. The causal effects of alcohol on human aggression were observed [106]. Acute alcohol intake is known to stimulates both dopamine and serotonin release. Serotonin depletion is linked to the rise in aggressive behavior [107] and alcohol intake in hostile individuals has an additive effect on aggression [108]. Wide literature supports the proposed circuit properties and suggests that active avoidance, anger and fear expression via amygdalar circuit is not only attenuated by vACC projections, but also by serotonin. After alcohol intake, the amygdala in persons with lower 5-HT signaling gets into over-drive by DA stimulation, causing more hostile reactions to threat. High DA levels weaken the inhibitory self-control in the APC, lead to underestimation of risks in dACC and increase impulsivity via D1 loop of NAc shell.

## Conclusions.

We linked the brain regions, which detect, react to, learn and predict what is bad or harmful and when things go wrong, processing information on threats to our well-being or survival. We proposed a causal role of this APC in the inhibitory self-control and learning to deselect – avoid, inhibit, stop going for the bad or suboptimal choices. The role of the APC is to guide our decisions and behavior away from bad consequences and from predictions, decisions, strategies or behavior, which went or were proved wrong. We claim that the APC competes with the reward processing circuit to bias the goal-directed behavior. They reciprocally attenuate each other's output: by direct projections, by opposing control of dopamine and serotonin signaling (via their opposite effects on the VTA/SNc, DRN and LHb) and by mostly opposite effects of dopamine and serotonin on stimulation versus inhibition of their regions. The net effect of the APC activation is antagonistic to the RPC activation. Both act in parallel to influence our thoughts, feelings and behavior ("don't do it" versus "go for it"). We showed DA and 5-HT effects on value-based control of behavior, motivation, decision making, moods, personality, well-being. We proposed main roles of dopamine and serotonin signaling in affective processing and choice behavior, and their dysfunctions in mental disorders.

We provided evidence for functional properties of the APC, which signals when we are not doing well. The adversities are identified by the AI, clOFC, pgACC and dACC. The information on bad or suboptimal events, choices or outcomes is then forwarded to the LHb, to inhibit dopamine and serotonin release via RMTg. We showed how the evaluations and interpretations in the AI, clOFC, dACC, mOFC and vACC predict affective meanings in the world, what is bad or good, and guide our thoughts and behavior. We predicted an affective bias in the cortical projections to the VS, with the APC preferentially linked to the D2 loop for inhibitory avoidance, while the RPC to the motivational D1 loop for wanting, seeking and moving for the choices. The good, rewarded choices guide us to go for them, by disinhibiting their MDT representations via GPi inhibition. The bad, punished choices get eventually avoided, by inhibition of MDT via GPi activation. Because the LHb reciprocally inhibits dopamine release in the VTA/SNc and serotonin release in the DRN, it stops us choosing and doing things linked to bad consequences but also makes us feel down, low, depressed when overstimulated.

We proposed that role of dACC is to detect when we are not doing well and to generate a warning to the real or predicted danger or risk - to prevent pain, harm, loss, failure or other bad consequences and suboptimal outcomes. It prods us to re-think our goals, improve our performance and adjust behavior after negative feedbacks/outcomes. This top-down warning signal from dACC calls for an update in our predictions, decisions and plans via lateral BA 10, enforces inhibitory avoidance via D2 loop of VS, corrects via preSMA or stops via STN the wrong motor responses, commands for cautious performance and focused attention via frontal eye fields (FEF) and intraparietal sulcus (IPS), for speed, alertness and alarm via NA, Ach and sympathetic control. The AI forms a functional module with the adjacent clOFC. Their proposed role is to detect something bad, low or wrong (noxious, harmful, unpleasant, faulty, false, strange, odd) in the quality of objects, persons, social conduct, information and concepts, to target those things, which should rather be avoided, as they might make us feel sick or unwell. AI/clOFC react with aversion, discomfort to bad, inferior, corrupted qualities and unacceptable behaviors, ideas.

We proposed and tested by wide literature data that DA generally attenuates (via their high D1/D2 receptors ratio on the GABAergic or low on pyramidal cells) the output of the APC. So it diminishes the warning function, inhibitory avoidance learning and self-control. On the contrary, DA enforces the output of the competing RPC, thus also the selection, drive, expended effort, preference and inclination to pursue choices and goals via the D1 loop of VS. We suggested that the reward value in mOFC, informational novelty in lateral BA 10 and informational value in the cognitive PFC induce dopamine release in the VTA, medial SNc and SNc, respectively. We claim that DA signals the expected value and meaning (good, right, new) to direct and move us towards the valuable, useful, important - survival and well-being promoting stuff and goals. It makes us motivated (even addicted), interested, biased, to choose, want, seek, move, work for the valuable choices, those which are reinforced by DA.

We came up with a new idea on the vACC role in the regulation of motivated behavior after gaining biological or social rewards. We proposed that vACC signals when we are doing well, when we gain or achieve good, valuable, well-being promoting outcomes: safety, resources, nurture, comfort, care, affection, acceptance, both when we reach goals by hard work or luck. We claim that via induction of serotonin in the DRN, vACC increases satisfaction, well-being, feeling alright, and decreases motivation, drive, wanting, appetite, impulsivity, violence, anxiety, worry, discomfort, pain, it slows us down and promotes rest state. In addition to its soothing effect on pain pathway, we propose that serotonin attenuates the dACC, AI/clOFC and LHb region of the APC, lowering discomfort, deprivation and anxiety. It calms down impulsivity, motivation and drive, via its inhibition of the D1 loop of VS, counteracting the dopamine effect on driving our behavior, by putting the brakes on wanting.

We proposed a calming, inhibitory effect of serotonin and potentiating effect of dopamine on the active avoidance and threat reactivity in the amygdalar circuit, in accordance with literature. The amygdala gets the input about threats from the same AI/clOFC and dACC regions that feed the inhibitory avoidance circuit. Contrary to the inhibitory avoidance, which is weakened by dopamine, the active avoidance is enforced by it. The active avoidance circuit involves the amygdala projections to D1 loop of VS, to BNST, PAG and PVN, from BNST to VTA and from VTA to VS. Its activation leads to active response to threat or stress, by self-defense, anger and aggression.

Our circuit based model adds new ideas on causes and consequences of affective, value-based processing and how they direct our behavior. The model is simple but its predictions are supported by wide literature on addiction, depression, anxiety, impulsivity, aggression, anhedonia, bipolar disorder, schizophrenia. The proposed region-specific effects of DA and 5-HT agents are useful for designing targeted drugs, to minimize their side effects. Our prediction of the value-based bias in cortical input to VS is testable using optogenetics. We suggest that the relative strength of dACC vs. vACC activation is an indicator of danger, alarm, things going wrong versus safety, well-being and things going well. Their ratio of activation can be used as marker in testing the efficiency of antidepressant drugs.